\documentclass[a4paper,fleqn,usenatbib]{mnras}

\usepackage[T1]{fontenc}
\usepackage{ae,aecompl}

\usepackage{graphicx}	
\usepackage{amsmath}	
\usepackage{amssymb}	

\title[Radio structure of J1100+4421]{The radio structure of the peculiar narrow-line Seyfert 1 galaxy candidate J1100+4421}

\author[K. \'E. Gab\'anyi et al.]{K. \'E. Gab\'anyi,$^{1,2}$\thanks{E-mail: gabanyi@konkoly.hu}
S. Frey,$^{1}$
Z. Paragi,$^{3}$
E. J\"arvel\"a,$^{4,5}$
T. Morokuma,$^{6}$ 
T. An,$^{7,8}$ \newauthor
M. Tanaka,$^{9}$
I. Tar$^{10}$ 
\\
$^{1}$Konkoly Observatory, MTA Research Centre for Astronomy and Earth Sciences, Konkoly Thege Mikl\'os \'ut 15-17, H-1121 Budapest, Hungary\\
$^{2}$MTA-ELTE Extragalactic Astrophysics Research Group, ELTE TTK P\'azm\'any P\'eter s\'et\'any 1/A, H-1117, Budapest, Hungary\\
$^{3}$Joint Institute for VLBI ERIC, P.O. Box 2, 7990 AA Dwingeloo, The Netherlands\\
$^{4}$Aalto University Mets\"ahovi Radio Observatory, Mets\"ahovintie 114, FI-02540 Kylm\"al\"a, Finland\\
$^{5}$Aalto University Department of Electronics and Nanoengineering, P.O. Box 15500, FI-00076 Aalto, Finland\\
$^{6}$Institute of Astronomy, Graduate School of Science, University of Tokyo, 2-21-1, Osawa, Mitaka, Tokyo 181-0015, Japan\\
$^{7}$Shanghai Astronomical Observatory, Chinese Academy of Sciences, 80 Nandan Road, 200030 Shanghai, P. R. China\\
$^{8}$Key Laboratory of Radio Astronomy, Chinese Academy of Sciences, 210008 Nanjing, P. R. China\\
$^{9}$National Astronomical Observatory of Japan, Mitaka, Tokyo 181-8588, Japan\\
$^{10}$Department of Astronomy, E\"otv\"os University, P.O. Box 32, H-1518 Budapest, Hungary\\
}

\date{Accepted XXX. Received YYY; in original form ZZZ}

\pubyear{2017}

\begin{document}
\label{firstpage}
\pagerange{\pageref{firstpage}--\pageref{lastpage}}
\maketitle

\begin{abstract}
Narrow-line Seyfert 1 galaxies (NLS1) are an intriguing subclass of active galactic nuclei. Their observed properties indicate low central black hole mass and high accretion rate. The extremely radio-loud NLS1 sources often show relativistic beaming and are usually regarded as younger counterparts of blazars. Recently, the object SDSS\,J110006.07$+$442144.3 was reported as a candidate NLS1 source. The characteristics of its dramatic optical flare indicated its jet-related origin. The spectral energy distribution of the object was similar to that of the $\gamma$-ray detected radio-loud NLS1, PMN\,J0948$+$0022. Our high-resolution European Very Long Baseline Interferometry Network observations at $1.7$ and $5$\,GHz revealed a compact core feature with a brightness temperature of $\gtrsim 10^{10}$\,K. 
Using the lowest brightness temperature value and assuming a moderate Lorentz factor of $\sim 9$ the jet viewing angle is $\lesssim 26^\circ$.
Archival Very Large Array data show a large-scale radio structure with a projected linear size of $\sim 150$\,kpc reminiscent of double-sided morphology. 
\end{abstract}




\section{Introduction}
Narrow-line Seyfert 1 galaxies \citep[NLS1, e.g.][]{Pogge2000} form a special subclass of active galactic nuclei (AGN). They were first recognized as Seyfert 1 galaxies with unusually narrow \ion{H}{i} lines \citep{OP1985}. NLS1 sources are identified by three features: (i) narrow permitted optical lines, with full width at half-maximum FWHM(H$\beta)<2000$\,km\,s$^{-1}$
, (ii) a flux ratio of [\ion{O}{iii}]$\lambda$5007 to H$\beta$ smaller than 3, and (iii) the strong emission feature caused by \ion{Fe}{ii} multiplets. The latter indicates a direct view to the accretion disk, although very recently, based upon the spectroscopic studies of a large sample of NLS1 candidate sources, \cite{new_spectral_def} concluded that NLS1 sources do not necessarily have strong \ion{Fe}{II} emission lines.
The relatively narrow widths of the permitted lines are explained with the lower velocities of the clouds in the broad-line region orbiting a low-mass black hole, $10^6 - 10^8$ M$_\odot$ \citep{AGN_evo}. These low-mass black holes in NLS1 sources accrete at high rates, close to the Eddington limit \citep[e.g. ][and references therein]{accretion_rate}. 

Concerning the radio emission, statistical studies showed that only $\sim 7$ per cent of NLS1 sources are radio loud \citep[e.g.][]{Zhou2006, Komossa2006}. The radio-loudness is commonly assessed following the original prescription of \cite{Kellermann_RL}, using the ratio of the $6$\,cm radio flux density to the $4400${\AA} optical flux density ($R$), and sources are usually classified to be radio-loud if $R\gtrsim10$ \citep[e.g.,][]{Komossa2006}. Most of the radio-loud NLS1 (RLNLS1) sources have steep spectra and resemble the compact steep-spectrum sources \citep{Komossa2006}. However, the extremely radio-loud NLS1s \citep[$\sim 2.5$ per cent of NLS1 sources, with $R\ge100$;][]{Komossa2006} show blazar-like properties: flat radio spectrum, compact radio cores, substantial variability, high brightness temperatures, flat X-ray spectra, and blazar-like spectral energy distribution \citep[e.g.][]{Yuan_blazarlikeNLS1}. Thus, these objects are thought to possess relativistic jets which are seen at a small angle to the line of sight, similarly to blazars. This claim was further strengthened by the {\it Fermi} satellite discovery of high-energy emission from a handful of RLNLS1 sources \citep[e.g.][and references therein]{Yao_newNLS1,vassilis_phd, 2fermi_tentative}. Further support of their blazar-like nature came from the  Monitoring Of Jets in Active galactic nuclei with VLBA Experiments \citep[MOJAVE,][]{mojave} survey. Among the five {\it Fermi}-detected RLNLS1 sources included in their sample, three show jet components moving at superluminal speeds (\citealt{lister_13}; see also \citealt{Fuhrmann_VLBA}).

Thus, flat-spectrum RLNLS1 sources are thought to be similar to blazars, but residing (mostly) in spiral host galaxies with central black holes of lower masses \citep{Crenshaw_host}. A statistical study by \cite{Berton_parentpop} showed that the black hole mass distribution is the same for steep-spectrum and flat-spectrum RLNLS1 sources. The suggested scenario of \cite{Berton_parentpop} is that at larger inclination angle, instead of flat-spectrum RLNLS1 sources, one starts to see steep-spectrum RLNLS1 sources. At even larger inclination angles, the Doppler effect broadens the line in the disk-like shaped broad-line regions \citep[e.g.,][]{disk-shape-BLR} and a broad-line radio galaxy hosted by a disk galaxy can be observed. Finally, when looking through the obscuring torus, the objects are described as narrow-line radio galaxies hosted by disk galaxies. 

The multi-wavelength study of \cite{Jarvela} showed that the jet is the main source of the radio, optical and X-ray emission in RLNLS1 sources, while in radio-quiet NLS1 sources the infrared and radio emission mostly originate from star formation. The origin of infrared emission in RLNLS1 sources however is unclear, it is possible that star-formation and reradiated emission from the torus also contribute \citep{WISE_NLRS1,Jarvela}.

SDSS\,J110006.07$+$442144.3 (hereafter, J1100$+$4421) is a newly-discovered NLS1 candidate source. It was found in the Kiso Supernova Survey \citep{Morokuma_Kiso} by \cite{disc} thanks to its dramatic optical flare on 2014 February 23. 

Follow-up observations of \cite{disc} revealed that the FWHMs of the broad components of \ion{Mg}{II} and H$\beta$ lines in J1100$+$4421 are smaller than $2000\mathrm{\,km\,s}^{-1}$. The redshift of the source was measured to be $z=0.84$. It was classified as an NLS1 candidate. \cite{disc} calculated the black hole mass, $\sim 1.5 \times 10^{7}$ M$_\odot$, using the \ion{Mg}{ii} line, and $\sim 1.0 \times 10^{7}$ M$_\odot$ using the H$\beta$ line. They also calculated the bolometric luminosity ($6 \times 10^{44}$\,erg\,s$^{-1}$), which is 30 per cent of the Eddington luminosity. These values are consistent with the findings of \cite{Foschini_RLNLS1}, who analyzed the multi-wavelength properties of 42 RLNLS1s, and found Eddington ratios of 1 per cent to 49 per cent, and black hole masses in the range of $10^6-10^8$ M$_\odot$. However J1100$+$4421 did not show a \ion{Fe}{ii} bump, and the flux ratio of the [\ion{O}{iii}] line to the H$\beta$ is much larger ($\sim 5-9$) than the value used to define NLS1 sources. According to \cite{disc}, the former can be attributed to the flaring state of the source, which can cause the emergence of strong continuum emission. With respect to the latter, the luminosity of the [\ion{O}{iii}] line was shown to correlate with the radio power \citep{oiii}. Thus, \cite{disc} argue that the especially luminous line in J1100$+$4421 may be related to strong jet emission in the source.
The presence of jet is also indicated by the extreme radio-loudness of J1100$+$4421. The ratio of the $1.4$\,GHz flux density measured in the Faint Images of the Radio Sky at Twenty-Centimeters (FIRST) survey \citep{first} to the quiescent flux density measured at $4400$\AA\, is $\sim 3000$. When the optical flux density is measured at the highest point of the flaring state, this ratio is still $\sim 400$ \citep{disc}. 

To image the radio structure of the source at milliarcsecond (mas) resolution and ascertain whether it has a blazar-like jet emission, we observed J1100$+$4421 with very long baseline interferometry (VLBI) using the European VLBI Network (EVN) at 1.7 and 5\,GHz. We also searched for a possible $\gamma$-ray counterpart to J1100$+$4421 in the latest {\it Fermi}/Large Arae Telescope (LAT) data. In the following, we assume a flat $\Lambda$CDM cosmological model with $H_0=70$\,km\,s$^{-1}$\,Mpc$^{-1}$, $\Omega_\textrm{m}=0.27$, and $\Omega_{\Lambda}=0.73$. At the redshift of the source ($z=0.84$), $1$\arcsec angular size corresponds to $7.754$\,kpc projected linear size \citep{calculator}.

\section{Observations and Data Reduction}
\subsection{EVN Data}

The EVN observations of J1100$+$4421 took place on 2015 February 10 at 1.7 GHz and on 2015 March 24 at 5 GHz. At 1.7 GHz, the interferometric array consisted of seven antennas: Effelsberg (Germany), the Jodrell Bank Mark 2 telescope (the United Kingdom), Medicina (Italy), Onsala (Sweden), Toru\'n (Poland), the Westerbork Synthesis Radio Telescope (WSRT, the Netherlands), and Sheshan (China). At 5 GHz, Noto (Italy) and Yebes (Spain) were also added to the array. However, Medicina did not observe and the WSRT produced no useful data at the higher frequency. 

The observations were carried out in e-VLBI mode \citep{eEVN}. The signals received at the radio telescopes were transmitted over optical fiber networks directly to the central data processor for real-time correlation. The correlation with $2$ s integration time was done at the EVN software correlator \citep[SFXC,][]{soft_corr}, in the Joint Institute for VLBI ERIC (JIVE), Dwingeloo, the Netherlands. Eight intermediate frequency channels (IFs) were used in both polarizations at both bands. Each IF had a width of $16$\,MHz and was divided into 32 spectral channels. The total bandwidth was $256$\,MHz.
The observations were performed in phase-reference mode \citep{phase-ref}. The target and the phase calibrator were observed alternately, with $\sim 3.5$\,min spent on the target and $\sim 1$\,min spent on the calibrator. The phase-reference calibrator used at both frequencies was J1108$+$4330, separated by $\sim 1.7\degr$ from the target in the sky. Its coordinates are right ascension $\alpha_\textrm{cal}=11^\textrm{h}08^\textrm{m} 23\fs47694$ and declination $\delta_\textrm{cal}=+43\degr 30\arcmin 53\farcs6571$, the uncertainties are $0.2$ mas in both directions.\footnote{Data are obtained from \url{http://astrogeo.org} maintained by L. Petrov, rfc2015b solutions} Additionally, 4C39.25 was included as fringe-finder at both bands. The observations lasted for $4.5$\,h and $4.25$\,h, and on-source integration times were $164$\,min and $119$\,min, at $1.7$ and $5$\,GHz, respectively.

The data were reduced in the standard manner \citep{data_reduc} using the U.S. National Radio Astronomy Observatory (NRAO) Astronomical Image Processing System \citep[{\sc aips},][]{aips}. First, interferometric visibility amplitudes were calibrated using the gain curves and system temperature measurements obtained at the telescope sites. Then fringe-fitting was performed on the phase-calibrator and fringe-finder sources. Their visibility data were exported to be imaged with the {\sc difmap} program package \citep{difmap}. The hybrid mapping procedure was used with several cycles of {\sc clean}ing \citep{clean} and phase self-calibration. Gain correction factors were determined in {\sc difmap} and subsequently were used in {\sc aips} to scale the amplitudes. The average corrections factors did not exceed 10 per cent. 
Four-four channels, the most affected by the bandpass response function, were discarded at the beginning and end of each IF.
Then fringe-fitting on the calibrator was repeated in {\sc aips}, now taking into account its brightness distribution using its {\sc clean} component model obtained in {\sc difmap}. The derived solutions were interpolated and applied to the target source, J1100+4421. The hybrid mapping of the target was also performed in {\sc difmap}. No amplitude self-calibration was attempted on the source, and phase self-calibration was not done for Sheshan for time intervals shorter than $60$\,min at $1.7$\,GHz and $15$\,min at $5$\,GHz. 

Since J1100$+$4421 was sufficiently bright and compact, we also performed fringe-fitting directly on the target source at both frequencies. Then we imaged the source in the same way as described above. The images obtained (Fig. \ref{fig:Lband} and Fig. \ref{fig:Cband}) were in good agreement with those obtained by transferring the phase solutions of the calibrator. Comparing the peak brightness values in the two images in both bands, the coherence loss in phase-referencing was $1$ per cent at $1.7$\,GHz and $14$ per cent at $5$\,GHz.

\subsection{Archival VLA data}

J1100$+$4421 was also observed during a $40$-s snapshot at $8.4$\,GHz with the Very Large Array (VLA) in its most extended A configuration on 1995 August 15 (project code: AM484) within the Cosmic Lens All-Sky Survey \citep{class}. The data were downloaded from the NRAO archive\footnote{\url{http://archive.nrao.edu}} and analyzed following standard data reduction steps in {\sc aips}. The absolute flux density calibrator used in the project was 3C286. The calibrated visibility data were exported from {\sc aips}, imaging was done in {\sc difmap}. The resulting map is shown in Fig. \ref{fig:VLA_8}. The weights of the data points are set inversely proportional to the amplitude errors (natural weighting).

\subsection{\emph{Fermi} Large Area Telescope data}

J1100+4421 has no associated counterpart in the newest and so far most comprehensive $\gamma$-ray catalogue the \emph{Fermi}/LAT 4-Year Point Source Catalogue \citep[3FGL, ][]{Fermi-cat4}. The closest 3FGL source is 3FGL J1105.7+4427 with a separation of more than $60 \arcmin$.

We analysed data from LAT using the \emph{Fermi}/LAT \texttt{Science Tools v10r0p5}{\footnote{\url{http://fermi.gsfc.nasa.gov/ssc/data/analysis/software/}}} software package. The LAT data were extracted in a circle with $15\degr$ radius around the coordinates of J1100+4421, over the whole energy range, choosing only LAT data type `photon'. We used the unbinned likelihood analysis and followed the provided tutorial{\footnote{\url{http://fermi.gsfc.nasa.gov/ssc/data/analysis/scitools/likelihood$\_$tutorial.html}}} in the data reduction. The parameters used for different tasks were the ones suggested in the tutorial. First, only events of class `source' in the energy band of 100 MeV -- 100 GeV were chosen using \texttt{gtselect}, and a zenith angle cut of $90\degr$ was implemented to remove the disturbance caused by the Earth's limb. Next \texttt{gtmktime} was used to select only the good time intervals for further analysis. Exposure map was generated using \texttt{gtltcube} and \texttt{gtexpmap}. The newest galactic diffuse emission model (gll$\_$iem$\_$v06.fits) and the extragalactic isotropic diffuse emission model (iso$\_$P8R2$\_$SOURCE$\_$V6$\_$v06.txt), which includes the residual cosmic-ray background, were used for computing the diffuse source responses with \texttt{gtdiffrps}. Finally, J1100+4421 and all the sources within $10\degr$ radius (29 sources) were modeled with \texttt{gtlike} using \texttt{powerlaw2} model and the unbinned likelihood algorithm with optimizer \texttt{MINUIT}.

We first integrated over the whole time period from 2008 August 5 00:00:00 UTC to 2016 December 19 00:00:00 UTC (3058 days) for which we have data. There are some very faint sources nearby, having test statistic (TS) values less than 0, causing the fit not to converge. We kept removing the sources with TS$<$0, except J1100+0044, and running the reduction again, until the TS values of all nearby sources were positive. Over the whole period the TS value for J1100+0044 was negative, meaning that the fit without the source is better than with it, implying that it is very improbable that this source has emitted detectable $\gamma$-rays during this period. To verify this result we integrated over the same time period in bins of 180 days, resulting in 17 bins, using the same parameters and procedure. The TS value for J1100+4421 was negative in every bin, further indicating that J1100+4421 has not been active at $\gamma$-rays between 2008 August and 2016 December. The upper limit of the $\gamma$-ray flux is $9.9 \times 10^{-10}$\,photon\,cm$^{-2}$\,s$^{-1}$ in the $100$\,MeV--$300$\,GeV energy band.

\section{Results}

\subsection{The pc-scale radio structure}

\begin{figure}
	\includegraphics[width=\columnwidth, bb=0 60 460 510, clip=true]{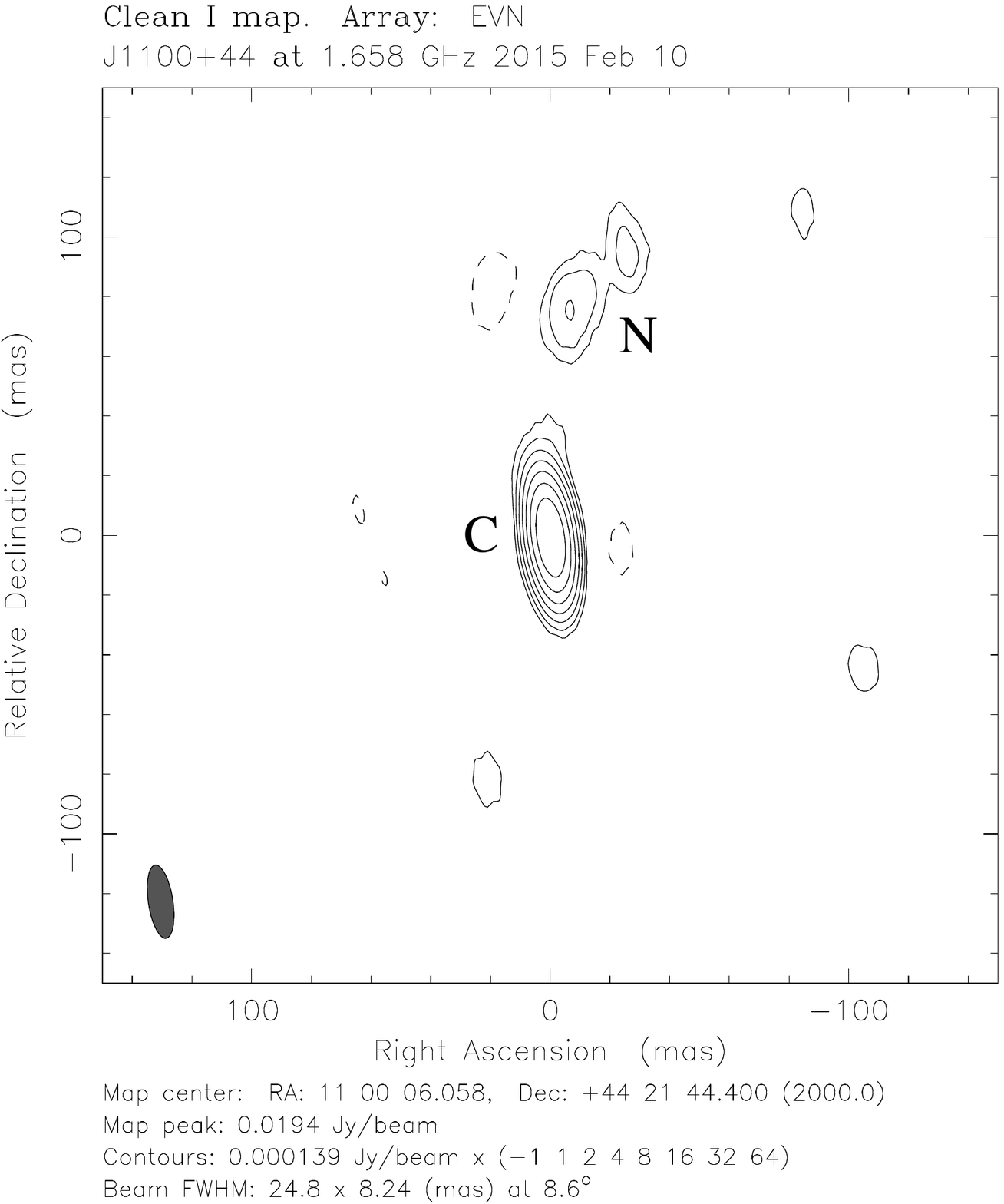}
    \caption{Naturally-weighted 1.7-GHz EVN image of J1100$+$4421. Observation took place on 2015 February 10. The peak brightness is $19.4$\,mJy\,beam$^{-1}$, the restoring beam is $24.8\textrm{\,mas} \times 8.2\textrm{\,mas}$ (FWHM) at a position angle of $9\degr$ and shown in the lower left corner of the image. The lowest contours are at $\pm 3.5\sigma$ noise level ($\pm 0.1$\,mJy\,beam$^{-1}$), further positive contour levels increase by a factor of two. The dashed line represents negative contour. The labels denote the Gaussian components fitted to the visibility data.}
    \label{fig:Lband}
\end{figure}

\begin{figure}
\includegraphics[width=\columnwidth, bb=0 60 410 510, clip=true]{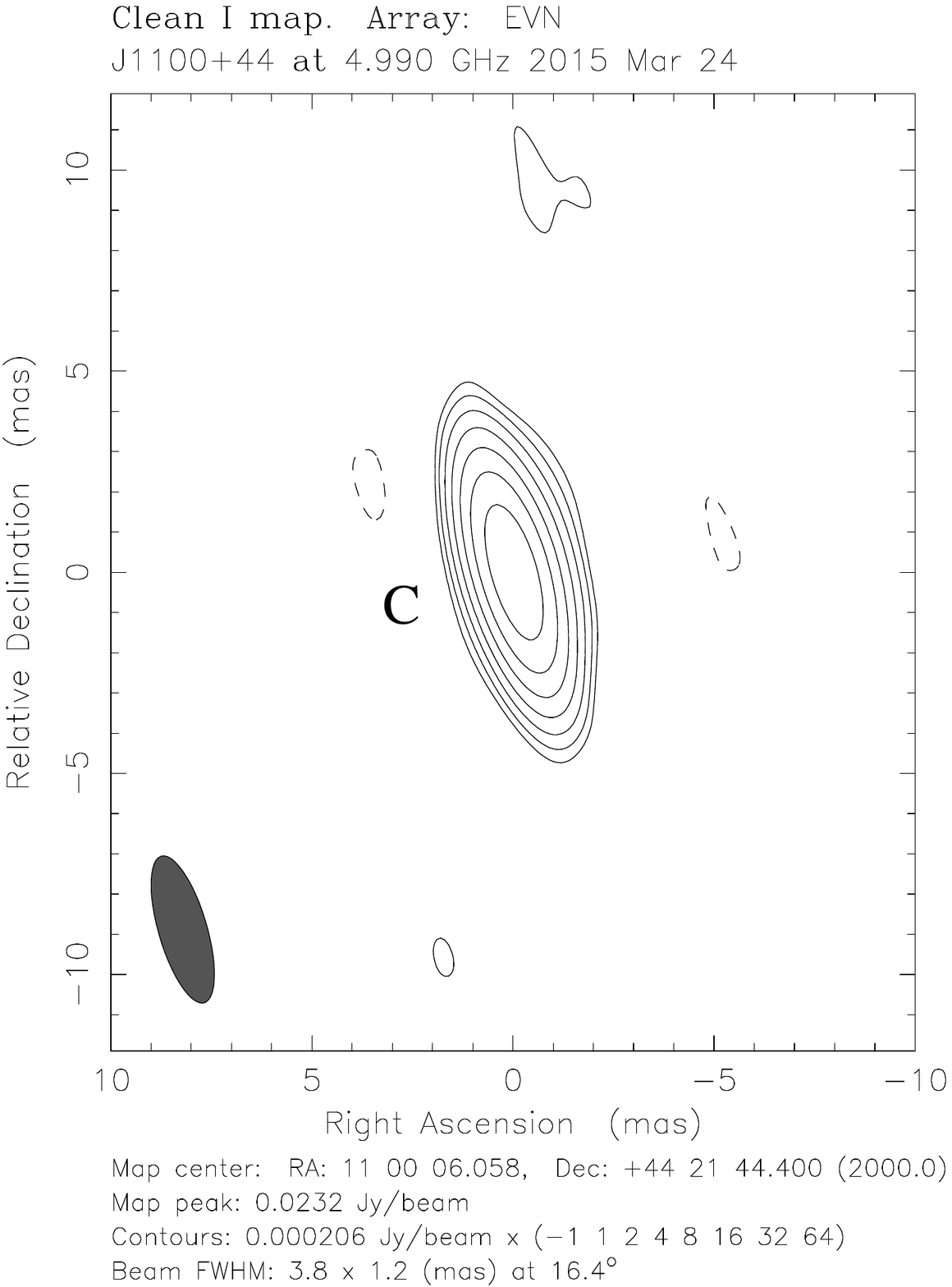}
    \caption{Naturally-weighted 5-GHz EVN image of J1100$+$4421. Observation took place on 2015 March 24. The peak brightness is $23.2$\,mJy\,beam$^{-1}$, the restoring beam is $3.8\textrm{\,mas} \times 1.2\textrm{\,mas}$ (FWHM) at a position angle of $16\degr$ and shown in the lower left corner of the image. The lowest contours are at $\pm 3.5\sigma$ noise level ($\pm  0.2 $\,mJy\,beam$^{-1}$), further positive contour levels increase by a factor of two. The dashed line represents negative contour. The label denotes the Gaussian component fitted to the visibility data.}
    \label{fig:Cband}
\end{figure}

A single, compact radio-emitting feature was detected at $5$\,GHz in our EVN observations (Fig. \ref{fig:Cband}). At $1.7$\,GHz, an additional feature is detected north from the central source (Fig. \ref{fig:Lband}). The phase-referencing observations allowed us to obtain the accurate coordinates of J1100$+$4421. We used the {\sc maxfit} verb in {\sc aips} to determine the right ascension and declination of the brightness peak at the $5$-GHz EVN image: $\alpha_\textrm{t}=11^\textrm{h}00^\textrm{m} 06\fs0571$, $\delta_\textrm{t}=+44\degr 21\arcmin 44\farcs383$. The coordinates are the same within the uncertainties at the two observing frequencies. We estimate that the obtained coordinates are accurate within $1$\,mas.

\begin{table*}
\caption{Details of the EVN observations of J1100$+$4421 (right ascension $\alpha_\textrm{t}=11^\textrm{h}00^\textrm{m} 06\fs0571$, declination $\delta_\textrm{t}=+44\degr 21\arcmin 44\farcs383$). In Cols. 2, 3, 4, 5, and 6, the parameters of the fitted circular Gaussian components are listed. In Cols. 7, 8, 9 and 10, the peak intensity, the size and position angle of the restoring beam and the rms noise levels of the radio images shown in Figs. \ref{fig:Lband} and \ref{fig:Cband} are given.}
\label{tab:radio}
\begin{tabular}{clcccccccc}
\hline
$\nu$ & ID & \multicolumn{2}{c}{Relative position} & Flux density & FWHM & peak & beam size & PA & rms \\
(GHz) & & RA (mas) & Dec. (mas) & (mJy) & (mas) & mJy\,beam$^{-1}$ & (mas$\times$mas) & (\degr) & (mJy\,beam$^{-1}$)\\
\hline
$1.7$ & C & -- & -- & $20.9 \pm 0.1$ & $2.5 \pm 0.2$ & 19.4 & $24.8\times 8.2$ & $9$ & $0.029$\\
1.7 & N & $ -12.5 \pm 2.0 $ & $79 \pm 2.0$ & $1.5 \pm 0.3$ & $16.0 \pm 4.0$ & -- & -- & -- & -- \\
$5$ & C & -- & -- & $25.0 \pm 0.4$ & $0.38 \pm 0.05$ & 23.2 & $3.8\times1.2$ & $16$& $0.057$\\
\hline
\end{tabular}
\end{table*}

We used {\sc difmap} to fit the visibilities with brightness distribution models. At $1.7$\,GHz, two circular Gaussian components are needed (labelled C and N), while at $5$\,GHz, a single circular Gaussian component (C) adequately modelled the visibilities. The parameters of the fitted components (projected separations from the core, flux densities and FWHM sizes) are summarized in Table \ref{tab:radio}.

Using the size and flux density values derived from the higher frequency measurement, one can calculate the brightness temperature as
\begin{equation}
T_\textrm{B}=1.22 \times 10^{12} \left(1+z\right) \frac{S}{\theta^2\nu^2}\,\mathrm{K} 
\end{equation}
where $z$ is the redshift of the source, $S$ is the flux density given in Jy, $\theta$ is the FWHM size in mas, and $\nu$ is the observing frequency in GHz. The obtained brightness temperature is $(1.56 \pm 0.4) \times 10^{10}$\,K. However, if the FWHM size derived from the modelfit is smaller than the smallest resolvable size of the array, the component is unresolved, and the brightness temperature value is only a lower limit. According to \cite{kovalev}, the minimum resolvable FWHM size of a Gaussian component in our observation at $5$\,GHz would be $0.18$\,mas along the major axis of the restoring beam, indicating that the feature is not unresolved. However, the more recent work of \cite{size_limit} showed that calibration uncertainties have significant effect on the minimum resolvable size. Investigating synthetic observations made with an array very similar to our EVN observation, they concluded that the minimum resolvable size can be a factor of two larger, $\sim 0.4$\,mas. Therefore, the fitted FWHM size of the component is an upper limit, and the derived brightness temperature of J1100$+$4421 can be regarded as a lower limit only.

The recovered flux density of the core component was somewhat larger at $5$\,GHz than at $1.7$\,GHz. Formally the compact source has a spectral index of $\alpha=0.20 \pm 0.03$ ($S \sim \nu^\alpha$). However, we note that the EVN observations were not simultaneous but separated by 6 weeks. Compact radio-emitting AGN are known to show variations in the radio on timescales of weeks to months.\footnote{Notably, near-infrared and optical monitoring of J1100$+$4421 by \cite{opt_nir} showed that J1100$+$4421 underwent a major flare a few days prior to our $5$\,GHz EVN observation.} Therefore, flux density variability may affect the spectral index estimate. 
We did not see at $5$\,GHz any counterpart of the northern feature detected at $1.7$\,GHz (Fig.\ref{fig:Lband}). The largest recovarable size of an interferometer array can be estimated as $\sim \lambda / (2 B_\mathrm{min})$, where $\lambda$ is the observing wavelength and $B_\mathrm{min}$ is the shortest baseline in the array \citep{las}. In our $5$\,GHz EVN observation, the shortest baseline is $B_\mathrm{min}\approx 637$\,km (the distance between Onsala and Toru\'n), thus the largest recoverable size is $9.7$\,mas. Therefore, a radio feature as large as the northern component in the $1.7$-GHz image, $16$\,mas, cannot be detected by our array used at $5$\,GHz. If it is smaller than the largest recoverable size ($9.7$\,mas) at $5$\,GHz, then its non-detection implies that its flux density is $\lesssim 2.1$\,mJy considering the $3.5\sigma$ image noise level ($0.2$\,mJy\,beam$^{-1}$). Either way, we cannot meaningfully constrain the spectral index of this component. However, its characteristics measured at $1.7$\,GHz are indicative of a more extended jet-related emission. The non-detection at $5$\,GHz also shows that there is no compact bright radio emitting region within this structure.

\subsection{The kpc-scale radio structure} \label{sect:res_kpc}

\begin{figure}
\includegraphics[width=\columnwidth, bb=42 180 565 600, clip=true]{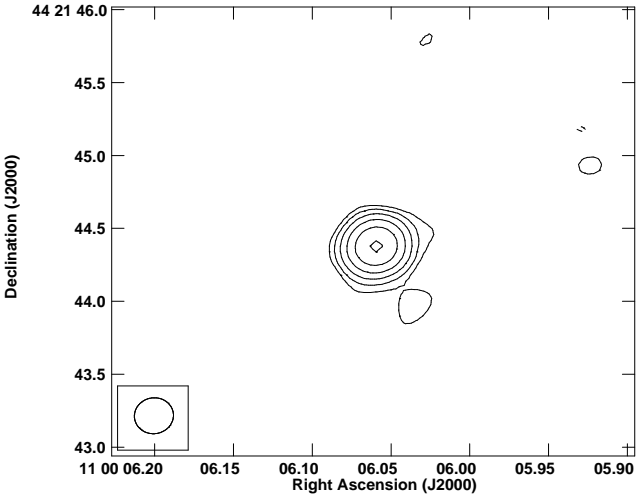}
    \caption{$8.4$-GHz VLA image of J1100$+$4421. The observation was performed on 1995 August 15. The peak brightness is $18.4$\,mJy\,beam$^{-1}$. The FWHM size of the restoring beam is $0\farcs27 \times 0\farcs25$ at a position angle of $-81\degr$, as shown in the bottom left corner of the image. The lowest contours are at $\pm 0.5$\,mJy\,beam$^{-1}$ corresponding to $3\sigma$ image noise level, further positive contour levels increase by a factor of two.}
    \label{fig:VLA_8}
\end{figure}

\begin{figure}
	\includegraphics[width=\columnwidth, bb=50 220 565 565, clip=]{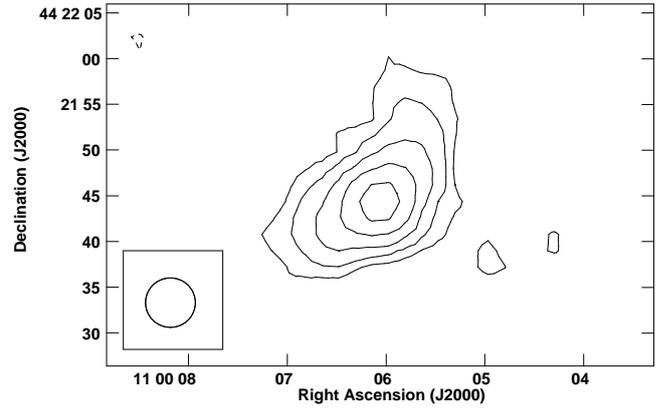}
    \caption{1.4-GHz VLA map of J1100$+$4421 from the FIRST survey \citep{first}. The observation was performed on 1997 Februray 25. The peak brightness is $8.3$\,mJy\,beam$^{-1}$, the restoring beam is circular with FWHM size of $5\farcs4$, and shown in the lower lef corner of the image. The lowest contour levels are at $\pm 0.4$\,mJy\,beam$^{-1}$ at $3\sigma$ image noise level, further positive contour levels increase by a factor of two. The dashed line represents negative contour.}
    \label{fig:FIRST}
\end{figure}


In the archival $8.4$-GHz VLA map of J1100$+$4421 (Fig. \ref{fig:VLA_8}) a single radio source can be seen. Model-fitting to the visibility data was performed in {\sc difmap}. A model containing a single circular Gaussian component adequately describes the data. Its flux density and FWHM size are $19.4 \pm 0.7$\,mJy and  $54.6 \pm 0.7\textrm{\,mas}$, respectively.

According to the FIRST \citep{first} survey, J1100$+$4421 shows an extended radio structure at $1.4$\,GHz (Fig. \ref{fig:FIRST}). In the most recent FIRST catalog \citep{Helfand_first}, the source is described with an elliptical Gaussian component with a FWHM size of $(7\farcs1 \pm 0\farcs3) \times (3\farcs0 \pm 0\farcs3)$, at a position angle of $131\degr \pm 2\degr$ and integral flux density of $15.76 \pm 0.4$\,mJy. (The errors of the flux density and the position angle are not given in the catalog, therefore we determined it from the FIRST image directly.) The peak brightness in the 1.4-GHz FIRST image is $8.3 \pm 0.1$\,mJy\,beam$^{-1}$. The mean epoch of the observation is $1997.158$ with an rms of $1.59$\,d. The flux density of the radio features detected with our EVN observation ($22.4$\,mJy) at a close frequency ($1.7$\,GHz) originates from the central region not resolved by FIRST, while the EVN is not sensitive to arcsec-scale emission recovered in FIRST. Thus the radio core of J1100$+$4421 must have brightened significantly during the $\sim 18$\,yr elapsed between the FIRST and EVN observations. 

At larger scales, J1100$+$4421 is unresolved in the 1.4-GHz NRAO VLA Sky Survey \citep[NVSS,][]{nvss}, with a flux density of $20.8 \pm 0.8$\,mJy. This is higher than the flux density measured in the FIRST survey. The difference can be because of possible additional large-scale structure that was resolved out in the FIRST map but was imaged at the lower resolution of the NVSS, or this can be also due to flux density variability of the compact core of J1100$+$4421. The NVSS observations took place two years before the FIRST observation. The mean epoch of the NVSS observations is $1995.25$, with an rms of $27.9$\,d. 

\section{Discussion}

\subsection{The characteristics of the VLBI jet}

The derived brightness temperature value is close to but does not exceed the commonly-used equipartition brightness temperature limit of $\sim 5 \times 10^{10}$\,K by \cite{readhead}. It is still lower than the intrinsic brightness temperature of $T_\mathrm{B,int}=3 \times 10^{10}$\,K derived by \cite{Homan_TB} for a sample of radio-loud AGN during their quiescent (not flaring) state, and thus the Doppler factor would be $\delta=T_\mathrm{B}/T_\mathrm{B,int}<1$. However, since the derived brightness temperature is just a lower limit, Doppler boosting in J1100$+$4421 cannot be excluded. If the source is indeed unresolved in our 5GHz EVN observation and the FWHM size of the emitting region is much smaller than the modelfit result, $<0.2$\,mas, the brightness temperature would exceed the equipartition limit.

In general the low brightness temperature value is not uncommon in RLNLS1 sources. \cite{Gu_VLBI} observed fourteen RLNLS1 sources with very high radio-loudness value ($R>100$) using the Very Long Baseline Array (VLBA) between $3.9$ and $7.9$\,GHz. They found that the brightness temperature values of the core components are within $10^{8.4}$\,K and $10^{11.4}$\,K, with a median value of $10^{10.1}$\,K, which agrees well with the brightness temperature we derived for J1100$+$4421. 
\cite{Gu_VLBI} explain the low brightness temperature values with intrinsically low jet power. Based upon the moderate radio variability seen in a sample of RLNLS1 sources, \cite{angelakis_nls1} also propose that the jets of RLNLS1 are mildly relativistic, and similar conclusion was drawn by \cite{richards-kpc} who investigated three RLNLS1 with large scale radio structures. It was hypothised that the lower bulk jet speed can be related to the lower black hole mass compared to the more energetic blazars \citep{Foschini_RLNLS1}.

If we assume a Lorentz factor e.g., $\gamma \sim 9$ which was derived by \cite{vassilis_phd} for the RLNLS1 source 1H\,0323$+$342 with a similar black hole mass \citep{Zhou_0323} as that of J1100$+$4421, and supposing that the brightness temperature is close to the derived lower limit, we obtain a viewing angle of $\sim 26\degr$. This should be considered as an upper limit on the actual viewing angle. The compact mas-scale radio morphology does indicate that the viewing angle of the jet cannot be too large, since above a few tens of degrees, one would expect to see a morphology reminiscent of a more resolved source rather than a single compact feature.

\cite{disc} pointed out that the observed optical variability is similar to that of the $\gamma$-ray detected RLNLS1 sources. They compared the spectral energy distribution of J1100$+$4421 to the $\gamma$-ray-loud RLNLS1s of PMN\,J0948$+$0022, and PKS\,2004$-$447 \citep{J0948_SED, three_gammarayNLS1} and hypothesized that assuming the same photon index as that of PMN\,J0948$+$0022, the upper limit of its $\gamma$-ray luminosity would be comparable to the measured $\gamma$-ray luminosity of PMN\,J0948$+$0022. However, our analysis of all available \emph{Fermi}/LAT data resulted in no detection of $\gamma$-rays at the position of J1100$+$4421. 

\cite{opt_nir} conducted extensive monitoring of J1100$+$4421 in the optical and near-infrared bands. They showed that based upon the spectral shape, the optical and near-infrared emission of the source is dominated by emission from the jet. From the fastest observed variations, and assuming a Doppler factor of $10$, they estimate that the region responsible for the variability must be smaller than $5.4 \times 10^{15}$\,cm. 


\subsection{The large-scale radio structure}

\begin{figure}
\includegraphics[width=\columnwidth, bb=25 10 730 508, clip=]{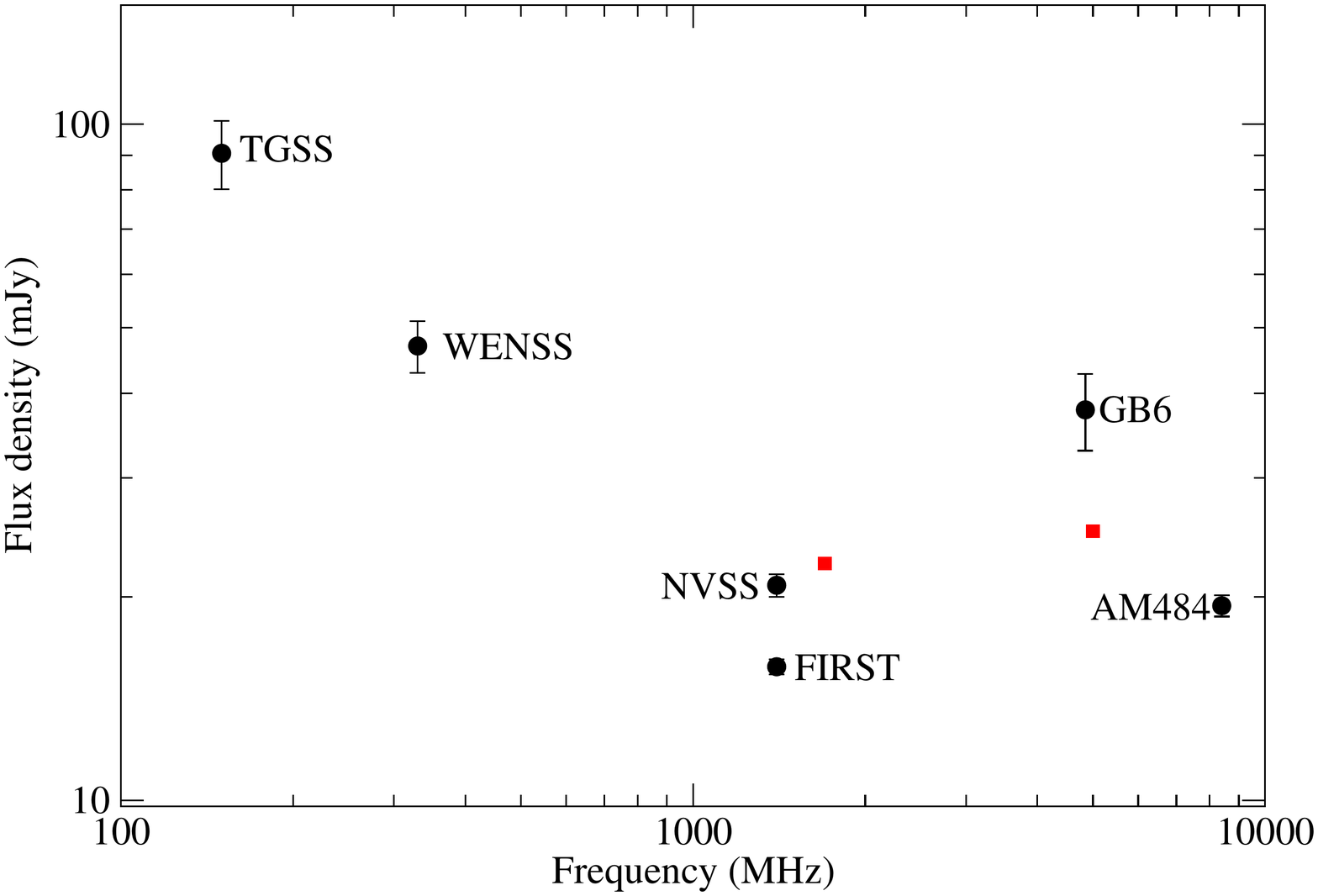}
    \caption{Non-contemporaneous radio flux density measurements of J1100$+$4421. Black circles show data from various surveys, the Giant Meterwave Radio Telescope 150 MHz all-sky radio survey \citep[TGSS,][]{TGSS}, the Westerbork Northern Sky Survey \citep[WENSS,][]{WENSS}, the NVSS, the FIRST, the Green Bank 4.85 GHz survey \citep[GB6,][]{GB6}, and the flux density from the archival 8.4-GHz VLA observation (project code: AM484). Red squares show the sum of the flux density of the fitted components of our EVN data.}   \label{fig:spectrum}
\end{figure}

In Fig. \ref{fig:spectrum}, the radio flux density measurements of J1100$+$4421 are summarized. The data are not simultaneous, and the observations are conducted with different angular resolutions, therefore source variability as well as resolution issues may complicate the picture. Nevertheless, up until 1.4\,GHz, the spectrum seems to follow a steep power-law, which might flatten afterwards. The red symbols represent the total flux density recovered in our high resolution EVN observations. The 1.7-GHz datapoint clearly shows the source variability (as discussed in Sect. \ref{sect:res_kpc}), while the discrepancy between the two 5-GHz data points can also be caused by the different resolutions of the observations (the EVN observation recovered only the pc-scale core emission, while in the Green Bank 4.85-GHz radio survey \citep{GB6} the kpc-scale scale structure could have been measured as well), as well as source variability.

The kpc-scale radio emission around J1100+4421 can only be seen in the FIRST image. In the 8-GHz VLA image, there was no additional radio emission above $0.8$\,mJy\,beam$^{-1}$, $4.5\sigma$ image noise level in a $30\arcsec \times 30\arcsec$ region around the source. This may indicate that the extended emission has a steep spectrum between $1.4$\,GHz and $8.4$\,GHz. The circular Gaussian component describing the northern feature detected at 1.7 GHz in our EVN observation is at $\sim 80$\,mas, corresponding to $\sim 620$\,pc projected distance form the core. This feature may be the jet connecting to the northern lobe seen in the FIRST image. There is no sign of any other radio emission down to a noise level of $0.13$\,mJy\,beam$^{-1}$ ($4\sigma$) at a region of $8\arcsec \times 8\arcsec$ around the center in the EVN image. If the two lobes contained compact features, the sensitivity of our EVN observation would have been adequate to detect them. Therefore, they are not compact, their sizes have to be larger than the largest recoverable size of the interferometer \citep{las}, which  is $\sim 70$\,mas in the 1.7-GHz EVN observation, corresponding to $\sim 540$\,pc at the redshift of J1100$+$4421.

The morphology of the large-scale structure of J1100$+$4421 can be classified as two-sided. \cite{doi-kpc} calculated the core dominance parameter for six RLNLS1 with kpc-scale radio structures as the flux density ratio between the core and lobes measured in the FIRST images. To disentangle the emission from the lobe and core region, we fitted the central region of the FIRST image of J1100$+$4421 with a circular Gaussian component. To keep this feature compact, we used a fixed size of $5\farcs4 \times 5\farcs4$, the clean beam size of the image. The resulting flux density of this unresolved component is $11.0\pm 0.2$\,mJy. Then we used the {\sc aips} verb {\sc imstat} to measure the flux density of the lobes in the residual image. They each have flux densities of $\sim 3$\,mJy. Thus, the core dominance parameter is $1.8$ for J1100$+$4421, the smallest among the core-dominated sources in \cite{doi-kpc}. Following \cite{doi-kpc}, we can estimate the jet kinetic energy in the large-scale radio structure of J1100$+$4421. The $1.4$-GHz radio power of a $\sim 3$\,mJy flux density lobe in J1100$+$4421 assuming a spectral index of $-0.7$ is $\sim 9 \times 10^{24} \mathrm{\,W\,Hz}^{-1}$. According to the scaling relation of \cite{jetpower} the jet kinetic power is thus $3 \times 10^{37}$\,W. This value is slightly higher than the minimum value, $\sim 10^{37}$\,W estimated to be necessary to create supersonic lobes according to \cite{doi-kpc}. Thus the radio power in the extended structure of J1100$+$4421 is intermediate between the radio powers in FRI and FRII radio galaxies \citep{FR}.

Currently, there are a handful RLNLS1 sources reported to have extended, kpc-scale radio structure \citep{whalen-kpc, anton-kpc, gliozzi-kpc, doi-kpc, richards-kpc, Congiu}, the projected size ranging between $\sim 23$\,kpc and $\sim 100$\,kpc. Additionally, recently \cite{Yao_newNLS1} reclassified a flat-spectrum quasar, SDSS\,J122222.55$+$041315.7 as an NLS1 source. According to its FIRST image and the VLA measurement at $1.4$\,GHz reported by \cite{new_largescale}, this RLNLS1 also exhibits a significant extended radio emission, with a projected linear size of $\sim 160$\,kpc. Thus the extended radio emission seen in J1100$+$4421 (Fig. \ref{fig:FIRST}), with a projected size of $\sim150$\,kpc at $1.4$\,GHz, is among the largest of the known RLNLS1 sources.

Assuming similar kpc-scale velocities as the ones for the six RLNLS1 sources investigated by \cite{doi-kpc}, it would require $\sim 10^6$\,yr to build the radio morphology seen in the FIRST image. However this is a lower limit only, since the viewing angle of the structure is not known, therefore its deprojected size might be larger. On the other hand, the similar flux density values seen in both lobes may indicate that the structure is oriented close to the plane of the sky.

\cite{opt_nir} report on the detection of a faint galaxy at a distance of $2\farcs7$ to the east from J1100$+$4421. We did not detect any radio counterpart at the position of this galaxy down to $0.5 \mathrm{\,mJy\,beam}^{-1}$ and $0.1 \mathrm{\,mJy\,beam}^{-1}$, $3 \sigma$ level from the VLA $8$-GHz data and our EVN $1.7$-GHz data, respectively. However it seems unlikely, it cannot be completely ruled out that the large-scale radio structure detected in the FIRST observation is partly related to this galaxy.

\section{Summary}

The source J1100$+$4421 was discovered during a dramatic optical flare by \cite{disc}. Based upon the follow-up optical observations, which showed unusually narrow broad lines, the source was classified as a candidate NLS1 source. However, the two other defining charateristics of NLS1 sources (presence of \ion{Fe}{II} bump and low flux of [\ion{O}{III}]$\lambda5007$ compared to H$\beta$ line) do not apply to J1100$+$4421. Nevertheless, the derived low black hole mass and relatively high Eddington rate are in agreement with the proposed possible NLS1 nature of the source \citep{new_spectral_def}. \cite{disc} suggested that the optical flare is most probably caused by the jet emission, and the optical and near-infrared monitoring campaign of \cite{opt_nir} showed that both the optical and near-infrared emissions are dominated by synchrotron radiation from the jet. 

Our high-resolution EVN observations showed that the source indeed has a mas-scale compact radio structure, similar to other very radio-loud NLS1 sources. The derived {\bf lower limit on the} brightness temperature is $1.56 \times 10^{10}$\,K, which is below but still consistent with the equipartition limit, therefore we cannot exclude the presence of Doppler boosting in the source. At lower resolution, J1100+4412 shows an extended, two-sided radio morphology of $\sim 150$\,kpc, which is among the largest of the known RLNLS1 sources. Somewhat contrary to the results of the VLBI observations, the large-scale radio emission indicates that the radio structure is oriented close to the plane of the sky. So far the source has not been detected in the {\it Fermi}/LAT $\gamma$-ray mission.

To reveal the connection and possible jet bending between the inner, compact radio-emitting feature and the large-scale lobes, high-sensivity, intermediate-resolution radio interferometric observations are desirable. Comparing the flux densities measured in different radio observations, it is clear that J1100$+$4421 shows significant variability at cm wavelengths at least on time scales of years. If the source show radio variability on shorter time scales, frequent radio flux density monitoring could provide independent estimate of the Doppler factor.

\section*{Acknowledgements}

We thank the referee for careful reading of the manuscript and for his/her comments, which helped to improve the paper. K. \'E. G. was supported by the J\'anos Bolyai Research Scholarship of the Hungarian Academy of Sciences. We wish to thank Venkatessh Ramakrishnan for the help with the {\it Fermi} data reduction.
The European VLBI Network is a joint facility of independent European, African, Asian, and North American radio astronomy institutes. Scientific results from data presented in this publication are derived from the following EVN project code: EG087.
The e-VLBI research infrastructure in Europe was supported by the European Community's Seventh Framework Programme (FP7/2007-2013) under grant agreement RI-261525 NEXPReS. The research leading to these results has received funding from the European Commission Seventh Framework Programme (FP/2007-2013) under grant agreement no. 283393 (RadioNet3). This research was supported by the Hungarian National Research Development and Innovation Office (OTKA NN110333) and the China--Hungary Collaboration and Exchange Programme by the International Cooperation Bureau of the Chinese Academy of Sciences. T. A. acknowledges the grant of the Youth Innovation Promotion Association of CAS.



\bibliographystyle{mnras}
\bibliography{ref} 

\begin{thebibliography}{}
\makeatletter
\relax
\def\mn@urlcharsother{\let\do\@makeother \do\$\do\&\do\#\do\^\do\_\do\%\do\~}
\def\mn@doi{\begingroup\mn@urlcharsother \@ifnextchar [ {\mn@doi@}
  {\mn@doi@[]}}
\def\mn@doi@[#1]#2{\def\@tempa{#1}\ifx\@tempa\@empty \href
  {http://dx.doi.org/#2} {doi:#2}\else \href {http://dx.doi.org/#2} {#1}\fi
  \endgroup}
\def\mn@eprint#1#2{\mn@eprint@#1:#2::\@nil}
\def\mn@eprint@arXiv#1{\href {http://arxiv.org/abs/#1} {{\tt arXiv:#1}}}
\def\mn@eprint@dblp#1{\href {http://dblp.uni-trier.de/rec/bibtex/#1.xml}
  {dblp:#1}}
\def\mn@eprint@#1:#2:#3:#4\@nil{\def\@tempa {#1}\def\@tempb {#2}\def\@tempc
  {#3}\ifx \@tempc \@empty \let \@tempc \@tempb \let \@tempb \@tempa \fi \ifx
  \@tempb \@empty \def\@tempb {arXiv}\fi \@ifundefined
  {mn@eprint@\@tempb}{\@tempb:\@tempc}{\expandafter \expandafter \csname
  mn@eprint@\@tempb\endcsname \expandafter{\@tempc}}}

\bibitem[\protect\citeauthoryear{{Abdo} et~al.,}{{Abdo}
  et~al.}{2009a}]{J0948_SED}
{Abdo} A.~A.,  et~al., 2009a, \mn@doi [\apj] {10.1088/0004-637X/699/2/976},
  \href {http://adsabs.harvard.edu/abs/2009ApJ...699..976A} {699, 976}

\bibitem[\protect\citeauthoryear{{Abdo} et~al.,}{{Abdo}
  et~al.}{2009b}]{three_gammarayNLS1}
{Abdo} A.~A.,  et~al., 2009b, \mn@doi [\apjl] {10.1088/0004-637X/707/2/L142},
  \href {http://adsabs.harvard.edu/abs/2009ApJ...707L.142A} {707, L142}

\bibitem[\protect\citeauthoryear{{Acero} et~al.,}{{Acero}
  et~al.}{2015}]{Fermi-cat4}
{Acero} F.,  et~al., 2015, \mn@doi [\apjs] {10.1088/0067-0049/218/2/23}, \href
  {http://adsabs.harvard.edu/abs/2015ApJS..218...23A} {218, 23}

\bibitem[\protect\citeauthoryear{{Angelakis} et~al.,}{{Angelakis}
  et~al.}{2015}]{angelakis_nls1}
{Angelakis} E.,  et~al., 2015, \mn@doi [\aap] {10.1051/0004-6361/201425081},
  \href {http://adsabs.harvard.edu/abs/2015A%26A...575A..55A} {575, A55}

\bibitem[\protect\citeauthoryear{{Ant{\'o}n}, {Browne}  \&
  {March{\~a}}}{{Ant{\'o}n} et~al.}{2008}]{anton-kpc}
{Ant{\'o}n} S.,  {Browne} I.~W.~A.,   {March{\~a}} M.~J.,  2008, \mn@doi [\aap]
  {10.1051/0004-6361:20078926}, \href
  {http://adsabs.harvard.edu/abs/2008A%26A...490..583A} {490, 583}

\bibitem[\protect\citeauthoryear{{Beasley} \& {Conway}}{{Beasley} \&
  {Conway}}{1995}]{phase-ref}
{Beasley} A.~J.,  {Conway} J.~E.,  1995, in {Zensus} J.~A.,  {Diamond} P.~J.,
  {Napier} P.~J.,  eds,  ASP Conf. Ser. Vol. 82, Very Long Baseline
  Interferometry and the VLBA. p.~327

\bibitem[\protect\citeauthoryear{{Becker}, {White}  \& {Helfand}}{{Becker}
  et~al.}{1995}]{first}
{Becker} R.~H.,  {White} R.~L.,   {Helfand} D.~J.,  1995, \mn@doi [ApJ]
  {10.1086/176166}, \href {http://adsabs.harvard.edu/abs/1995ApJ...450..559B}
  {450, 559}

\bibitem[\protect\citeauthoryear{{Berton} et~al.,}{{Berton}
  et~al.}{2015}]{Berton_parentpop}
{Berton} M.,  et~al., 2015, \mn@doi [\aap] {10.1051/0004-6361/201525691}, \href
  {http://adsabs.harvard.edu/abs/2015A%26A...578A..28B} {578, A28}

\bibitem[\protect\citeauthoryear{{Caccianiga} et~al.,}{{Caccianiga}
  et~al.}{2015}]{WISE_NLRS1}
{Caccianiga} A.,  et~al., 2015, \mn@doi [\mnras] {10.1093/mnras/stv939}, \href
  {http://adsabs.harvard.edu/abs/2015MNRAS.451.1795C} {451, 1795}

\bibitem[\protect\citeauthoryear{{Cavagnolo}, {McNamara}, {Nulsen}, {Carilli},
  {Jones}  \& {B{\^i}rzan}}{{Cavagnolo} et~al.}{2010}]{jetpower}
{Cavagnolo} K.~W.,  {McNamara} B.~R.,  {Nulsen} P.~E.~J.,  {Carilli} C.~L.,
  {Jones} C.,   {B{\^i}rzan} L.,  2010, \mn@doi [\apj]
  {10.1088/0004-637X/720/2/1066}, \href
  {http://adsabs.harvard.edu/abs/2010ApJ...720.1066C} {720, 1066}

\bibitem[\protect\citeauthoryear{{Collin} \& {Kawaguchi}}{{Collin} \&
  {Kawaguchi}}{2004}]{accretion_rate}
{Collin} S.,  {Kawaguchi} T.,  2004, \mn@doi [\aap]
  {10.1051/0004-6361:20040528}, \href
  {http://adsabs.harvard.edu/abs/2004A%26A...426..797C} {426, 797}

\bibitem[\protect\citeauthoryear{{Condon}, {Cotton}, {Greisen}, {Yin},
  {Perley}, {Taylor}  \& {Broderick}}{{Condon} et~al.}{1998}]{nvss}
{Condon} J.~J.,  {Cotton} W.~D.,  {Greisen} E.~W.,  {Yin} Q.~F.,  {Perley}
  R.~A.,  {Taylor} G.~B.,   {Broderick} J.~J.,  1998, \mn@doi [AJ]
  {10.1086/300337}, \href {http://adsabs.harvard.edu/abs/1998AJ....115.1693C}
  {115, 1693}

\bibitem[\protect\citeauthoryear{{Congiu} et~al.,}{{Congiu}
  et~al.}{2017}]{Congiu}
{Congiu} E.,  et~al., 2017, \mn@doi [\aap] {10.1051/0004-6361/201730616}, \href
  {http://adsabs.harvard.edu/abs/2017A%26A...603A..32C} {603, A32}

\bibitem[\protect\citeauthoryear{{Cracco}, {Ciroi}, {Berton}, {Di Mille},
  {Foschini}, {La Mura}  \& {Rafanelli}}{{Cracco}
  et~al.}{2016}]{new_spectral_def}
{Cracco} V.,  {Ciroi} S.,  {Berton} M.,  {Di Mille} F.,  {Foschini} L.,  {La
  Mura} G.,   {Rafanelli} P.,  2016, \mn@doi [\mnras] {10.1093/mnras/stw1689},
  \href {http://adsabs.harvard.edu/abs/2016MNRAS.462.1256C} {462, 1256}

\bibitem[\protect\citeauthoryear{{Crenshaw}, {Kraemer}  \& {Gabel}}{{Crenshaw}
  et~al.}{2003}]{Crenshaw_host}
{Crenshaw} D.~M.,  {Kraemer} S.~B.,   {Gabel} J.~R.,  2003, \mn@doi [\aj]
  {10.1086/377625}, \href {http://adsabs.harvard.edu/abs/2003AJ....126.1690C}
  {126, 1690}

\bibitem[\protect\citeauthoryear{{Diamond}}{{Diamond}}{1995}]{data_reduc}
{Diamond} P.~J.,  1995, in {Zensus} J.~A.,  {Diamond} P.~J.,   {Napier} P.~J.,
  eds,  ASP Conf. Ser. Vol. 82, Very Long Baseline Interferometry and the VLBA.
  Astron. Soc. Pac., San Francisco, p.~227

\bibitem[\protect\citeauthoryear{{Doi}, {Nagira}, {Kawakatu}, {Kino}, {Nagai}
  \& {Asada}}{{Doi} et~al.}{2012}]{doi-kpc}
{Doi} A.,  {Nagira} H.,  {Kawakatu} N.,  {Kino} M.,  {Nagai} H.,   {Asada} K.,
  2012, \mn@doi [\apj] {10.1088/0004-637X/760/1/41}, \href
  {http://adsabs.harvard.edu/abs/2012ApJ...760...41D} {760, 41}

\bibitem[\protect\citeauthoryear{{Fanaroff} \& {Riley}}{{Fanaroff} \&
  {Riley}}{1974}]{FR}
{Fanaroff} B.~L.,  {Riley} J.~M.,  1974, \mn@doi [\mnras]
  {10.1093/mnras/167.1.31P}, \href
  {http://adsabs.harvard.edu/abs/1974MNRAS.167P..31F} {167, 31P}

\bibitem[\protect\citeauthoryear{{Foschini}}{{Foschini}}{2011}]{2fermi_tentative}
{Foschini} L.,  2011, in {Foschini} L.,  {Colpi} M.,  {Gallo} L.,  {Grupe} D.,
  {Komossa} S.,  {Leighly} K.,   {Mathur} S.,  eds, Narrow-Line Seyfert 1
  Galaxies and their Place in the Universe. Proceedings of Science (PoS,
  Trieste, Italy), p.~24

\bibitem[\protect\citeauthoryear{{Foschini} et~al.,}{{Foschini}
  et~al.}{2015}]{Foschini_RLNLS1}
{Foschini} L.,  et~al., 2015, \mn@doi [\aap] {10.1051/0004-6361/201424972},
  \href {http://adsabs.harvard.edu/abs/2015A%26A...575A..13F} {575, A13}

\bibitem[\protect\citeauthoryear{{Fuhrmann} et~al.,}{{Fuhrmann}
  et~al.}{2016}]{Fuhrmann_VLBA}
{Fuhrmann} L.,  et~al., 2016, \mn@doi [Research in Astronomy and Astrophysics]
  {10.1088/1674-4527/16/11/176}, \href
  {http://adsabs.harvard.edu/abs/2016RAA....16..176F} {16, 176}

\bibitem[\protect\citeauthoryear{{Gliozzi}, {Papadakis}, {Grupe}, {Brinkmann},
  {Raeth}  \& {Kedziora-Chudczer}}{{Gliozzi} et~al.}{2010}]{gliozzi-kpc}
{Gliozzi} M.,  {Papadakis} I.~E.,  {Grupe} D.,  {Brinkmann} W.~P.,  {Raeth} C.,
    {Kedziora-Chudczer} L.,  2010, \mn@doi [\apj]
  {10.1088/0004-637X/717/2/1243}, \href
  {http://adsabs.harvard.edu/abs/2010ApJ...717.1243G} {717, 1243}

\bibitem[\protect\citeauthoryear{{Gregory}, {Scott}, {Douglas}  \&
  {Condon}}{{Gregory} et~al.}{1996}]{GB6}
{Gregory} P.~C.,  {Scott} W.~K.,  {Douglas} K.,   {Condon} J.~J.,  1996,
  \mn@doi [\apjs] {10.1086/192282}, \href
  {http://cdsads.u-strasbg.fr/abs/1996ApJS..103..427G} {103, 427}

\bibitem[\protect\citeauthoryear{{Greisen}}{{Greisen}}{2003}]{aips}
{Greisen} E.~W.,  2003, \mn@doi [Information Handling in Astronomy - Historical
  Vistas] {10.1007/0-306-48080-8_7}, \href
  {http://adsabs.harvard.edu/abs/2003ASSL..285..109G} {285, 109}

\bibitem[\protect\citeauthoryear{{Gu}, {Chen}, {Komossa}, {Yuan}, {Shen},
  {Wajima}, {Zhou}  \& {Zensus}}{{Gu} et~al.}{2015}]{Gu_VLBI}
{Gu} M.,  {Chen} Y.,  {Komossa} S.,  {Yuan} W.,  {Shen} Z.,  {Wajima} K.,
  {Zhou} H.,   {Zensus} J.~A.,  2015, \mn@doi [\apjs]
  {10.1088/0067-0049/221/1/3}, \href
  {http://adsabs.harvard.edu/abs/2015ApJS..221....3G} {221, 3}

\bibitem[\protect\citeauthoryear{{Helfand}, {White}  \& {Becker}}{{Helfand}
  et~al.}{2015}]{Helfand_first}
{Helfand} D.~J.,  {White} R.~L.,   {Becker} R.~H.,  2015, \mn@doi [\apj]
  {10.1088/0004-637X/801/1/26}, \href
  {http://adsabs.harvard.edu/abs/2015ApJ...801...26H} {801, 26}

\bibitem[\protect\citeauthoryear{{H\"ogbom}}{{H\"ogbom}}{1979}]{clean}
{H\"ogbom} J.~A.,  1979, A\&AS, \href
  {http://adsabs.harvard.edu/abs/1979A%26AS...36..173H} {36, 173}

\bibitem[\protect\citeauthoryear{{Homan} et~al.,}{{Homan}
  et~al.}{2006}]{Homan_TB}
{Homan} D.~C.,  et~al., 2006, \mn@doi [ApJ] {10.1086/504715}, \href
  {http://adsabs.harvard.edu/abs/2006ApJ...642L.115H} {642, L115}

\bibitem[\protect\citeauthoryear{{Intema}, {Jagannathan}, {Mooley}  \&
  {Frail}}{{Intema} et~al.}{2017}]{TGSS}
{Intema} H.~T.,  {Jagannathan} P.,  {Mooley} K.~P.,   {Frail} D.~A.,  2017,
  \mn@doi [\aap] {10.1051/0004-6361/201628536}, \href
  {http://cdsads.u-strasbg.fr/abs/2017A%26A...598A..78I} {598, A78}

\bibitem[\protect\citeauthoryear{{J{\"a}rvel{\"a}}, {L{\"a}hteenm{\"a}ki}  \&
  {Le{\'o}n-Tavares}}{{J{\"a}rvel{\"a}} et~al.}{2015}]{Jarvela}
{J{\"a}rvel{\"a}} E.,  {L{\"a}hteenm{\"a}ki} A.,   {Le{\'o}n-Tavares} J.,
  2015, \mn@doi [\aap] {10.1051/0004-6361/201424694}, \href
  {http://adsabs.harvard.edu/abs/2015A%26A...573A..76J} {573, A76}

\bibitem[\protect\citeauthoryear{{Karamanavis}}{{Karamanavis}}{2015}]{vassilis_phd}
{Karamanavis} V.,  2015, PhD thesis, Max-Planck-Institut f{\"u}r
  Radioastronomie, \mn@doi{10.5281/zenodo.48650}

\bibitem[\protect\citeauthoryear{{Keimpema} et~al.,}{{Keimpema}
  et~al.}{2015}]{soft_corr}
{Keimpema} A.,  et~al., 2015, \mn@doi [Experimental Astronomy]
  {10.1007/s10686-015-9446-1}, \href
  {http://adsabs.harvard.edu/abs/2015ExA....39..259K} {39, 259}

\bibitem[\protect\citeauthoryear{{Kellermann}, {Sramek}, {Schmidt}, {Shaffer }
  \& {Green}}{{Kellermann} et~al.}{1989}]{Kellermann_RL}
{Kellermann} K.~I.,  {Sramek} R.,  {Schmidt} M.,  {Shaffer } D.~B.,   {Green}
  R.,  1989, \mn@doi [\aj] {10.1086/115207}, \href
  {http://adsabs.harvard.edu/abs/1989AJ.....98.1195K} {98, 1195}

\bibitem[\protect\citeauthoryear{{Kharb}, {Lister}  \& {Cooper}}{{Kharb}
  et~al.}{2010}]{new_largescale}
{Kharb} P.,  {Lister} M.~L.,   {Cooper} N.~J.,  2010, \mn@doi [\apj]
  {10.1088/0004-637X/710/1/764}, \href
  {http://adsabs.harvard.edu/abs/2010ApJ...710..764K} {710, 764}

\bibitem[\protect\citeauthoryear{{Komossa}, {Voges}, {Xu}, {Mathur}, {Adorf},
  {Lemson}, {Duschl}  \& {Grupe}}{{Komossa} et~al.}{2006}]{Komossa2006}
{Komossa} S.,  {Voges} W.,  {Xu} D.,  {Mathur} S.,  {Adorf} H.-M.,  {Lemson}
  G.,  {Duschl} W.~J.,   {Grupe} D.,  2006, \mn@doi [\aj] {10.1086/505043},
  \href {http://adsabs.harvard.edu/abs/2006AJ....132..531K} {132, 531}

\bibitem[\protect\citeauthoryear{{Kovalev} et~al.,}{{Kovalev}
  et~al.}{2005}]{kovalev}
{Kovalev} Y.~Y.,  et~al., 2005, \mn@doi [AJ] {10.1086/497430}, \href
  {http://adsabs.harvard.edu/abs/2005AJ....130.2473K} {130, 2473}

\bibitem[\protect\citeauthoryear{{Labiano}}{{Labiano}}{2008}]{oiii}
{Labiano} A.,  2008, \mn@doi [\aap] {10.1051/0004-6361:200810399}, \href
  {http://adsabs.harvard.edu/abs/2008A%26A...488L..59L} {488, L59}

\bibitem[\protect\citeauthoryear{{Lister} et~al.,}{{Lister}
  et~al.}{2009}]{mojave}
{Lister} M.~L.,  et~al., 2009, \mn@doi [\aj] {10.1088/0004-6256/138/6/1874},
  \href {http://adsabs.harvard.edu/abs/2009AJ....138.1874L} {138, 1874}

\bibitem[\protect\citeauthoryear{{Lister} et~al.,}{{Lister}
  et~al.}{2016}]{lister_13}
{Lister} M.~L.,  et~al., 2016, \mn@doi [\aj] {10.3847/0004-6256/152/1/12},
  \href {http://adsabs.harvard.edu/abs/2016AJ....152...12L} {152, 12}

\bibitem[\protect\citeauthoryear{{Mathur}}{{Mathur}}{2000}]{AGN_evo}
{Mathur} S.,  2000, \mn@doi [\mnras] {10.1046/j.1365-8711.2000.03530.x}, \href
  {http://adsabs.harvard.edu/abs/2000MNRAS.314L..17M} {314, L17}

\bibitem[\protect\citeauthoryear{{Morokuma} et~al.,}{{Morokuma}
  et~al.}{2014}]{Morokuma_Kiso}
{Morokuma} T.,  et~al., 2014, \mn@doi [\pasj] {10.1093/pasj/psu105}, \href
  {http://adsabs.harvard.edu/abs/2014PASJ...66..114M} {66, 114}

\bibitem[\protect\citeauthoryear{{Morokuma} et~al.,}{{Morokuma}
  et~al.}{2017}]{opt_nir}
{Morokuma} T.,  et~al., 2017, preprint, \href
  {http://adsabs.harvard.edu/abs/2017arXiv170705416M} {} (\mn@eprint {arXiv}
  {1707.05416})

\bibitem[\protect\citeauthoryear{{Myers} et~al.,}{{Myers} et~al.}{2003}]{class}
{Myers} S.~T.,  et~al., 2003, \mn@doi [\mnras]
  {10.1046/j.1365-8711.2003.06256.x}, \href
  {http://cdsads.u-strasbg.fr/abs/2003MNRAS.341....1M} {341, 1}

\bibitem[\protect\citeauthoryear{{Natarajan}, {Paragi}, {Zwart}, {Perkins},
  {Smirnov}  \& {van der Heyden}}{{Natarajan} et~al.}{2017}]{size_limit}
{Natarajan} I.,  {Paragi} Z.,  {Zwart} J.,  {Perkins} S.,  {Smirnov} O.,   {van
  der Heyden} K.,  2017, \mn@doi [\mnras] {10.1093/mnras/stw2653}, \href
  {http://cdsads.u-strasbg.fr/abs/2017MNRAS.464.4306N} {464, 4306}

\bibitem[\protect\citeauthoryear{{Osterbrock} \& {Pogge}}{{Osterbrock} \&
  {Pogge}}{1985}]{OP1985}
{Osterbrock} D.~E.,  {Pogge} R.~W.,  1985, \mn@doi [\apj] {10.1086/163513},
  \href {http://adsabs.harvard.edu/abs/1985ApJ...297..166O} {297, 166}

\bibitem[\protect\citeauthoryear{{Pogge}}{{Pogge}}{2000}]{Pogge2000}
{Pogge} R.~W.,  2000, \mn@doi [\nar] {10.1016/S1387-6473(00)00065-8}, \href
  {http://adsabs.harvard.edu/abs/2000NewAR..44..381P} {44, 381}

\bibitem[\protect\citeauthoryear{{Readhead}}{{Readhead}}{1994}]{readhead}
{Readhead} A.~C.~S.,  1994, \mn@doi [ApJ] {10.1086/174038}, \href
  {http://adsabs.harvard.edu/abs/1994ApJ...426...51R} {426, 51}

\bibitem[\protect\citeauthoryear{{Rengelink}, {Tang}, {de Bruyn}, {Miley},
  {Bremer}, {Roettgering}  \& {Bremer}}{{Rengelink} et~al.}{1997}]{WENSS}
{Rengelink} R.~B.,  {Tang} Y.,  {de Bruyn} A.~G.,  {Miley} G.~K.,  {Bremer}
  M.~N.,  {Roettgering} H.~J.~A.,   {Bremer} M.~A.~R.,  1997, \mn@doi [\aaps]
  {10.1051/aas:1997358}, \href
  {http://adsabs.harvard.edu/abs/1997A%26AS..124..259R} {124}

\bibitem[\protect\citeauthoryear{{Richards} \& {Lister}}{{Richards} \&
  {Lister}}{2015}]{richards-kpc}
{Richards} J.~L.,  {Lister} M.~L.,  2015, \mn@doi [\apjl]
  {10.1088/2041-8205/800/1/L8}, \href
  {http://adsabs.harvard.edu/abs/2015ApJ...800L...8R} {800, L8}

\bibitem[\protect\citeauthoryear{{Shen} \& {Ho}}{{Shen} \&
  {Ho}}{2014}]{disk-shape-BLR}
{Shen} Y.,  {Ho} L.~C.,  2014, \mn@doi [\nat] {10.1038/nature13712}, \href
  {http://adsabs.harvard.edu/abs/2014Natur.513..210S} {513, 210}

\bibitem[\protect\citeauthoryear{{Shepherd}, {Pearson}  \& {Taylor}}{{Shepherd}
  et~al.}{1994}]{difmap}
{Shepherd} M.~C.,  {Pearson} T.~J.,   {Taylor} G.~B.,  1994, BAAS, \href
  {http://adsabs.harvard.edu/abs/1994BAAS...26..987S} {26, 987}

\bibitem[\protect\citeauthoryear{{Szomoru}}{{Szomoru}}{2008}]{eEVN}
{Szomoru} A.,  2008, in Proceedings of the 9th European VLBI Network Symposium
  on The role of VLBI in the Golden Age for Radio Astronomy and EVN Users
  Meeting. Proceedings of Science (PoS, Trieste, Italy), p.~40

\bibitem[\protect\citeauthoryear{{Tanaka} et~al.,}{{Tanaka}
  et~al.}{2014}]{disc}
{Tanaka} M.,  et~al., 2014, \mn@doi [\apjl] {10.1088/2041-8205/793/2/L26},
  \href {http://adsabs.harvard.edu/abs/2014ApJ...793L..26T} {793, L26}

\bibitem[\protect\citeauthoryear{{Whalen}, {Laurent-Muehleisen}, {Moran}  \&
  {Becker}}{{Whalen} et~al.}{2006}]{whalen-kpc}
{Whalen} D.~J.,  {Laurent-Muehleisen} S.~A.,  {Moran} E.~C.,   {Becker} R.~H.,
  2006, \mn@doi [\aj] {10.1086/500825}, \href
  {http://adsabs.harvard.edu/abs/2006AJ....131.1948W} {131, 1948}

\bibitem[\protect\citeauthoryear{{Wright}}{{Wright}}{2006}]{calculator}
{Wright} E.~L.,  2006, \mn@doi [\pasp] {10.1086/510102}, \href
  {http://adsabs.harvard.edu/abs/2006PASP..118.1711W} {118, 1711}

\bibitem[\protect\citeauthoryear{{Wrobel}}{{Wrobel}}{1995}]{las}
{Wrobel} J.~M.,  1995, in {Zensus} J.~A.,  {Diamond} P.~J.,   {Napier} P.~J.,
  eds,  ASP Conf. Ser. Vol. 82, Very Long Baseline Interferometry and the VLBA.
  p.~411

\bibitem[\protect\citeauthoryear{{Yao}, {Yuan}, {Zhou}, {Komossa}, {Zhang},
  {Qiao}  \& {Liu}}{{Yao} et~al.}{2015}]{Yao_newNLS1}
{Yao} S.,  {Yuan} W.,  {Zhou} H.,  {Komossa} S.,  {Zhang} J.,  {Qiao} E.,
  {Liu} B.,  2015, \mn@doi [\mnras] {10.1093/mnrasl/slv119}, \href
  {http://adsabs.harvard.edu/abs/2015MNRAS.454L..16Y} {454, L16}

\bibitem[\protect\citeauthoryear{{Yuan}, {Zhou}, {Komossa}, {Dong}, {Wang},
  {Lu}  \& {Bai}}{{Yuan} et~al.}{2008}]{Yuan_blazarlikeNLS1}
{Yuan} W.,  {Zhou} H.~Y.,  {Komossa} S.,  {Dong} X.~B.,  {Wang} T.~G.,  {Lu}
  H.~L.,   {Bai} J.~M.,  2008, \mn@doi [\apj] {10.1086/591046}, \href
  {http://adsabs.harvard.edu/abs/2008ApJ...685..801Y} {685, 801}

\bibitem[\protect\citeauthoryear{{Zhou}, {Wang}, {Yuan}, {Lu}, {Dong}, {Wang}
  \& {Lu}}{{Zhou} et~al.}{2006}]{Zhou2006}
{Zhou} H.,  {Wang} T.,  {Yuan} W.,  {Lu} H.,  {Dong} X.,  {Wang} J.,   {Lu} Y.,
   2006, \mn@doi [\apjs] {10.1086/504869}, \href
  {http://adsabs.harvard.edu/abs/2006ApJS..166..128Z} {166, 128}

\bibitem[\protect\citeauthoryear{{Zhou} et~al.,}{{Zhou}
  et~al.}{2007}]{Zhou_0323}
{Zhou} H.,  et~al., 2007, \mn@doi [\apjl] {10.1086/513604}, \href
  {http://adsabs.harvard.edu/abs/2007ApJ...658L..13Z} {658, L13}

\makeatother
\end{thebibliography}







\bsp	
\label{lastpage}
\end{document}